\documentclass[
  superscriptaddress,
  preprint,
  final,
  aps,
  prb,
  amsfonts,
  letterpaper,
  citeautoscript,
  draft,
  endfloats
]{revtex4-1}

\usepackage{graphicx}
\usepackage{multirow,bm, amsmath}
\graphicspath{{./figs/}}    
\usepackage{hyperref}
\usepackage{bbm}
\usepackage[usenames,dvipsnames]{color}
\usepackage{subfigure}

\renewcommand{\vec}[1]{\ensuremath\boldsymbol{#1}}
\newcommand{\ibraket}[2]{\bigl{<}#1\big|#2\bigr{>}}
\newcommand{\eV}{\text{eV}}
\newcommand{\A}{\text{A}}
\newcommand{\B}{\text{B}}
\providecommand{\C}{\text{C}}
\newcommand{\I}{\text{I}}

\newcommand{\fig}[1]{Fig.~\ref{#1}}
\newcommand{\Fig}[1]{Figure~\ref{#1}}
\newcommand{\eq}[1]{Eq.~(\ref{#1})}

\newlength{\myhgt}

\begin{document}


\title{Ab-initio prediction of fast non-equilibrium transport of nascent polarons in SrI$_2$: A key to high-performance scintillation}

\author{Fei Zhou}
\affiliation{Physical and Life Sciences Directorate, Lawrence Livermore National Laboratory, Livermore, California 94550, USA}
\author{Babak Sadigh}
\affiliation{Physical and Life Sciences Directorate, Lawrence Livermore National Laboratory, Livermore, California 94550, USA}
\author{Paul Erhart}
\affiliation{Chalmers University of Technology, Department of Physics, S-412 96 Gothenburg, Sweden}
\author{Daniel \AA{}berg}
\affiliation{Physical and Life Sciences Directorate, Lawrence Livermore National Laboratory, Livermore, California 94550, USA}

\date{\today} 

\begin{abstract}
The excellent light yield proportionality of europium-doped strontium iodide (SrI$_2$:Eu) has resulted in state-of-the-art $\gamma$-ray detectors with remarkably high energy resolution, far exceeding that of most halide compounds.
In this class of materials the formation of self-trapped hole polarons is very common. However, polaron formation is usually expected to limit carrier mobilities and has been associated with poor scintillator light-yield proportionality and resolution. 
Here, using a recently developed first-principles method, we perform an unprecedented study of polaron transport in SrI$_2$, both for equilibrium polarons, as well as nascent polarons immediately following a self-trapping event. We propose a rationale for the unexpected high energy resolution of SrI$_2$. We identify nine stable hole polaron configurations, which consist of dimerized iodine pairs with polaron binding energies of up to 0.5\,eV. They are connected by a complex potential energy landscape that comprises 66 unique nearest-neighbor migration paths. \textit{Ab initio} molecular dynamics simulations reveal that a large fraction of polarons is born into configurations that migrate practically barrier free at room temperature. Consequently, carriers created during $\gamma$-irradiation can quickly diffuse away reducing the chance for non-linear recombination, the primary culprit for non-proportionality and resolution reduction.
We conclude that the flat, albeit complex, landscape for polaron migration in SrI$_2$ is key for understanding its outstanding performance. This insight provides important guidance not only for the future development of high-performance scintillators but also of other materials, for which large polaron mobilities are crucial such as batteries and solid state ionic conductors.
\end{abstract}

\maketitle


\section*{Introduction}

A small, or self-trapped, polaron is a quasi-particle consisting of a localized charge carrier that has strongly polarized its immediate surrounding lattice. Polaron formation and transport are crucial for understanding important quantum-mechanical processes occurring in scintillations \cite{WilSon90}, as well as batteries \cite{MaxZhoCed06}, and solid ion conductors. The polaron can be centered on a specific atom, e.g., hole and electron polarons in TiO$_2$ or SrTiO$_3$ \cite{DeaAraFra12, Erhart2014PRB035204}, or atomic bonds, e.g., in the case of $V_K$-centers in alkali halides \cite{Kanzig1955PR1890}. The latter consists of a hole localized on a dimerized nearest-neighbor halide-ion pair. For small polarons such as these, the adiabatic potential energy landscape (APES) comprises potential wells that are usually less than 1\,eV deep (corresponding to the polaron binding energy) and separated by energy barriers that are generally lower than the binding energy. As a result, polaron migration typically proceeds by a hopping mechanism without delocalization.

Since standard density functional (DFT) theory fails to stabilize polaron states \cite{Gavartin2003PRB035108, SadErhAbe15}, mainly due to the large self-interaction error of the localized state, several alternatives have been suggested. These are either based on introducing a localized potential in the spirit of DFT+$U$ \cite{Anisimov1991PRB943, LanZun09, Erhart2014PRB035204} or, in the case of hybrid functionals, by incorporating a certain fraction of possibly screened exchange interaction \cite{Gavartin2003PRB035108, DeaAraFra12, BisDu12, SadErhAbe15}. Although the dependence on external parameters in both approaches can to some extent be removed by invoking Koopman's theorem, both approaches currently suffer from severe drawbacks. Approaches based on local potentials are commonly restricted to polarons localized on atomic sites and thus exclude $V_K$-centers in alkali halides. Hybrid functionals do not suffer from this shortcoming but are associated with a much larger computational cost. Nevertheless, migration barriers in highly symmetric systems such as sodium iodide have been studied through brute force hybrid functional calculations \cite{SadErhAbe15}. 
Beyond these very simple structures, where the local polaron geometry and migration pathways have been known for decades, the computational cost associated with sampling the APES of extended (periodic) systems has severely impacted theoretical studies.

In particular, these difficulties have been a major obstacle in understanding the polaronic properties of state-of-the art scintillator materials such as SrI$_2$ \cite{CheHulDro08, WilLoeGlo08, ChePayAsz09, AleKhoHaa12, ErhSchSad14}, LaBr$_3$ \cite{LoeDorEij01, AbeSadErh12}, or Cs$_2$LiYCl$_6$ \cite{PawSpa97, BesDorEij04, BisDu12}. These materials exhibit a much larger chemical, structural and electronic complexity than ``classic'' alkali halide materials, possibly leading to unusual effects on polaron formation and migration.

Here, we capitalize on a recently devised parameter-free energy functional, the polaron self-interaction correction (pSIC) method, that is suitable for studying polaronic properties and is computationally much more efficient than hybrid XC functionals (up to a factor of 500 for the cases in Ref.~\onlinecite{SadErhAbe15}). As a result, we are able to construct a comprehensive map of the APES for hole polarons in SrI$_2$. We show that the energy landscape features pathways that give rise to an unusually fast polaron migration compared to materials with similar polaron binding energies. This finding represents a critical step towards our understanding of the excellent scintillator performance of this material. More generally, it demonstrates a powerful approach for the exploration of polaron migration in complex materials.



\section*{Results}

\subsection{Stable polaron configurations}

$V_K$-centers in simple alkali halides form when the removal of an electron from the top of the valence band, which is dominated by halogen-$p$ states, creates an open-shell system. This triggers the formation of a covalent $5p\sigma$ bond between neighboring halogen ions and implies the ejection of an unoccupied antibonding $p\sigma^*$ state into the gap. 

While SrI$_2$ has a much more complex crystal structure than e.g., NaI, it is reasonable to expect a similar mechanism also in SrI$_2$, as the top of the valence band consists of iodine $5p$ states \cite{ErhSchSad14}.  Relaxations starting from randomly distorted structures as well as subsequent molecular dynamics simulations using pSIC indeed only find polaronic states of this kind. A random search is, however, not suitable to identify all possible dimer pairs in a system as complex as SrI$_2$ and hence a more systematic approach has to be employed. To this end, we considered all symmetrically distinct first-nearest neighbor iodine pairs between two I sites A and B in the ideal SrI$_2$ structure. 

\begin{figure}
  \includegraphics[ width=0.88\linewidth]{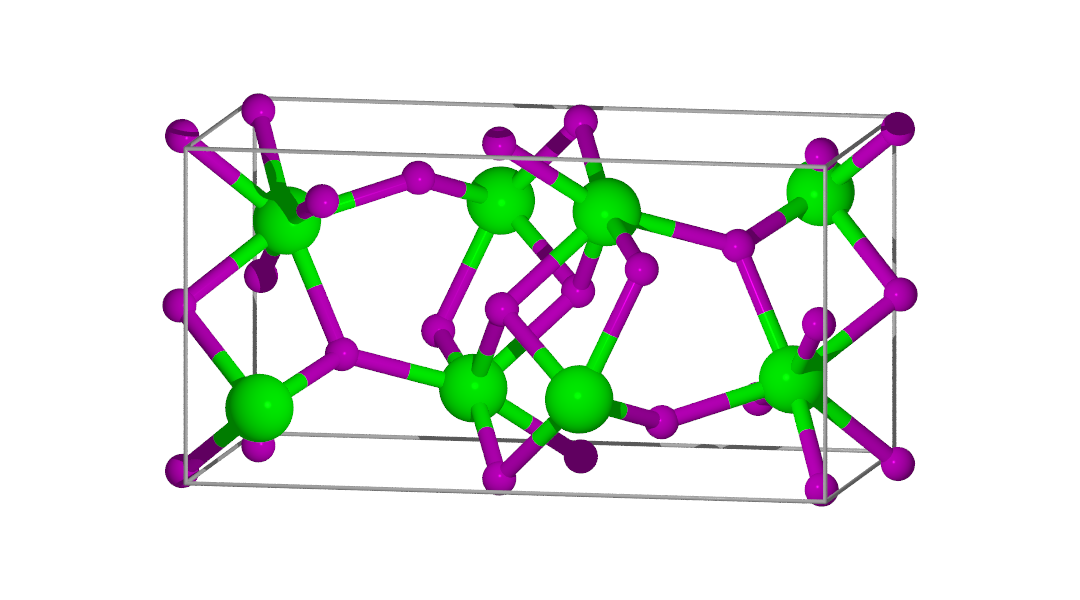}
  \includegraphics[ width=0.92\linewidth]{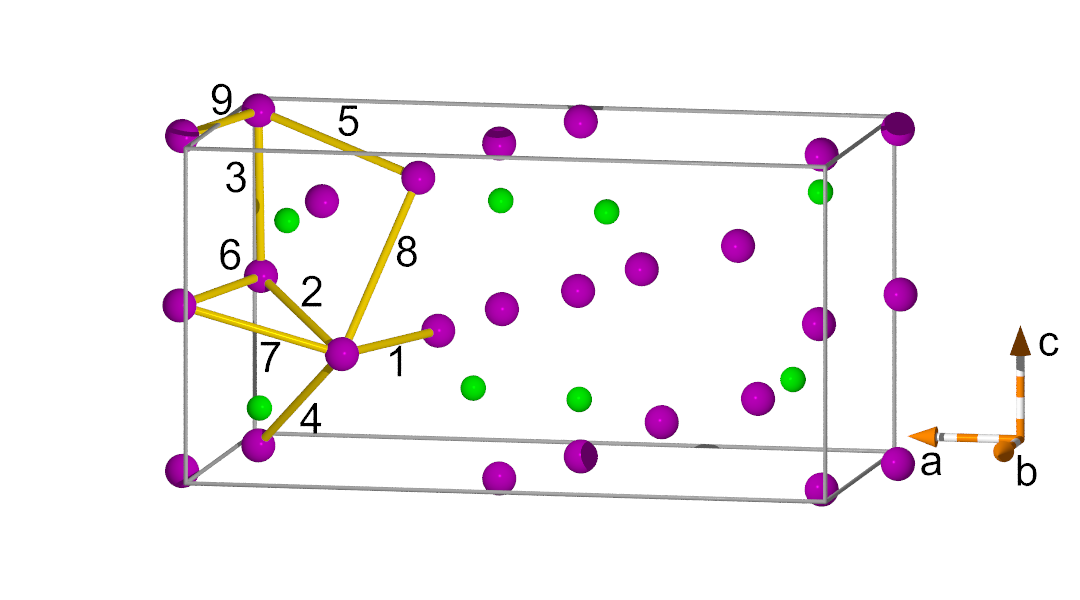}
  \caption{
    Crystal structure of SrI$_2$ and hole polaron configurations.
    (a) Perspective view of the primitive unit cell; (b) all stable I--I self-trapped polarons numbered 1 through 9, with only one representative dimer shown per type.
    Green and lila spheres represent Sr and I ions, respectively.
  }
  \label{fig:structure} 
\end{figure}

In the ideal SrI$_2$ structure [Fig.~\ref{fig:structure}(a)], there are 12 symmetrically distinct I--I pairs that are separated by 3.9--5.0\,\AA, while the second iodine coordination shell is more than 5.7\,\AA\ apart. Initial seed structures for polaron configurations were constructed from these pairs by displacing the two I ions toward each other to obtain a separation of 3\,\AA\ with all other atoms fixed. Nine of the possible first-nearest neighbor I--I pairs remain stable after pSIC relaxation with the Perdew-Becke-Ern\-zer\-hof (PBE) \cite{Perdew1996PRL3865} functional, all resulting in a dimer bond-lengths of approximately 3.3 \AA\ [Fig.~\ref{fig:structure}(b)], in accordance with recent results for $V_K$ centers in NaI and LiI \cite{SadErhAbe15}. The three remaining I--I pairs are unstable with respect to relaxation into one of the nine stable ones. Subsequent relaxations of the nine stable pSIC-relaxed structures with hybrid functional calculations using the PBE0 parametrization \cite{Adamo1999JCP6158} do not appreciably affect the geometry. 

\begin{figure}
  \includegraphics[width=\linewidth]{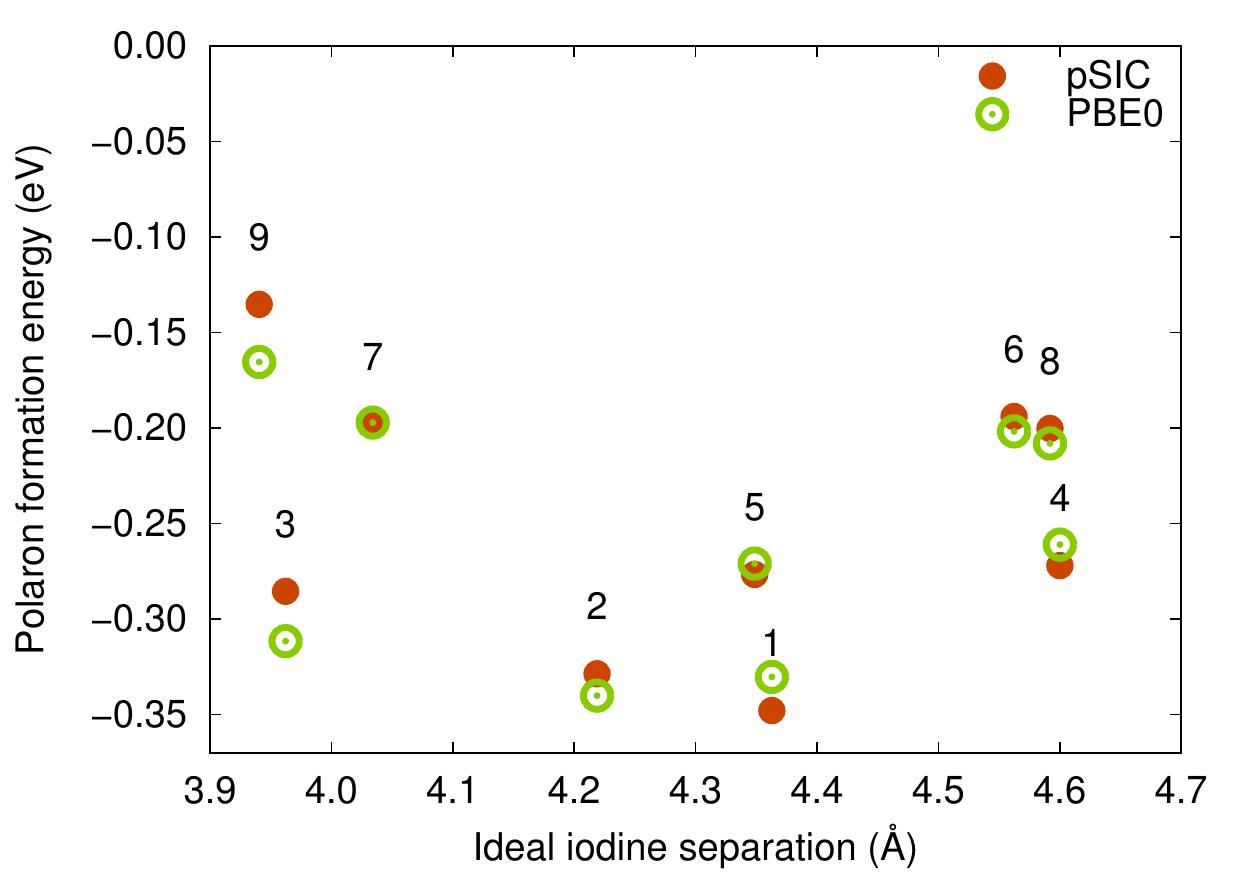}
  \caption{
    Polaron formation energy \textit{vs} the initial (ideal) I--I dimer separation distance from pSIC (filled circles) and PBE0 (squares) calculations obtained using 72-atom supercells. PBE0 values are image-charge corrected.
  }
  \label{fig:formation-separation}
\end{figure}

The formation energies obtained at the pSIC level range from $-0.34\,\eV$ (configuration \textbf{1}) to $-0.17\,\eV$ (configuration \textbf{9}) [Fig.~\ref{fig:formation-separation}]. The most stable dimers (\textbf{1} and \textbf{2}) originate from configurations with an initial I--I separation of 4.2--$4.4\,\text{\AA}$. This is somewhat contrary to the naive expectation that the dimer with the shortest ideal bond length would be energetically favorable. The formation energies obtained using the pSIC method compare very favorably with the results from the computationally more demanding hybrid PBE0 calculations [\fig{fig:formation-separation}].

\begin{figure}[htp]
  \includegraphics[width=\linewidth]{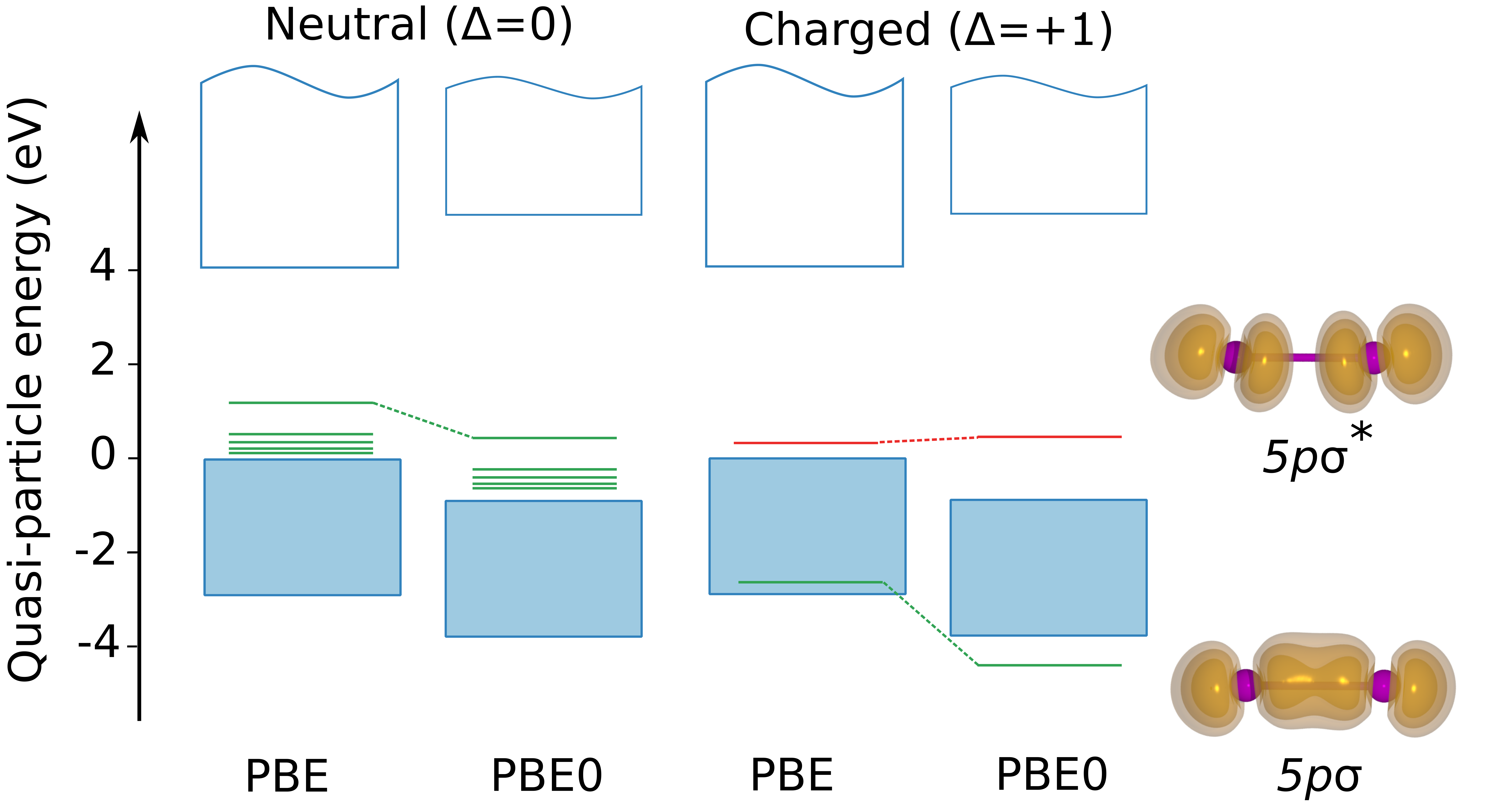}
  \caption{
    Energy levels of the neutral and charged cell of the most stable polaron configuration.
    In the neutral charge state PBE and PBE0 agree with respect to the relative position of the occupied $5p\sigma^*$ antibonding state.
    In the charged case, PBE, however, fails to describe the position of both the antibonding $5p\sigma^*$ and bonding $5p\sigma$ state.
  }
  \label{fig:energy-level}
\end{figure}

To further elucidate this relation, we considered the (quasiparticle) energy levels for PBE and PBE0 in both the neutral and charged state for the most energetically favorable $V_K$-center configuration (\fig{fig:energy-level}). In the charge neutral case for both PBE and PBE0, {\em five} occupied levels, all related to $5p$ states of the dimerized iodine pair, are ejected into the band gap, with the topmost occupied level being the antibonding $5p\sigma^*$ state. PBE and PBE0 closely agree with respect to the positions of the ejected levels relative to the valence band edge, and also the overlap $\ibraket{\psi_{\text{PBE}}}{\psi_{\text{PBE0}}}$ for the five ejected bands is very close to unity.

The charged state, however, highlights the failure of PBE to describe polaronic states. While the wave function overlap of the now unoccupied antibonding $5p\sigma^*$ state is still close to one,
the energetic position relative to the valence band edge differs significantly between PBE to PBE0. Similarly, the bonding $5p\sigma$ state resides {\em below} the valence band in PBE0 while it forms a resonance {\em inside} the valence band in the case of PBE.

\subsection{Polaron migration}
\label{sec:mig}

\begin{figure*}
    \subfigure{
      \includegraphics[width=0.46\textwidth]{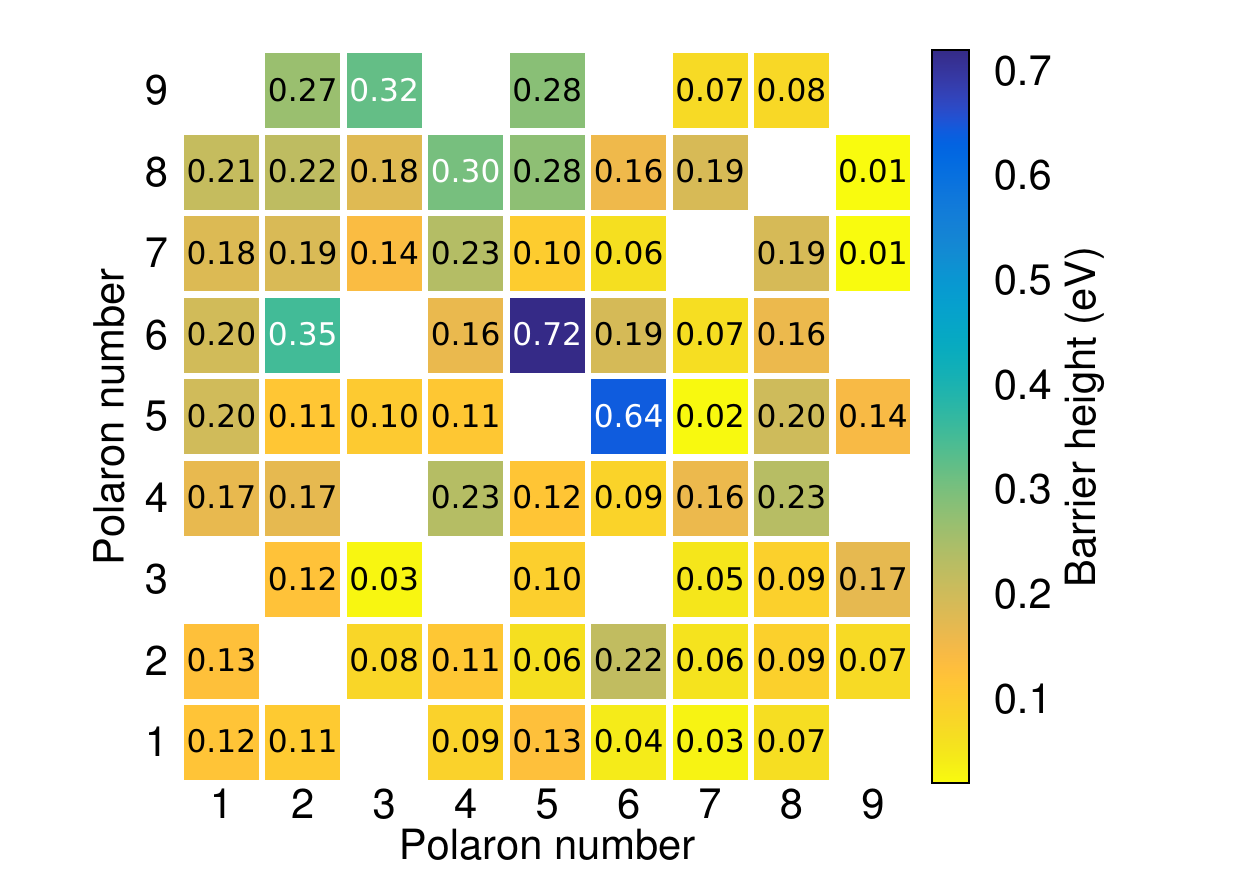}
      \label{fig:table}
}
\subfigure{
          \includegraphics[height=\myhgt]{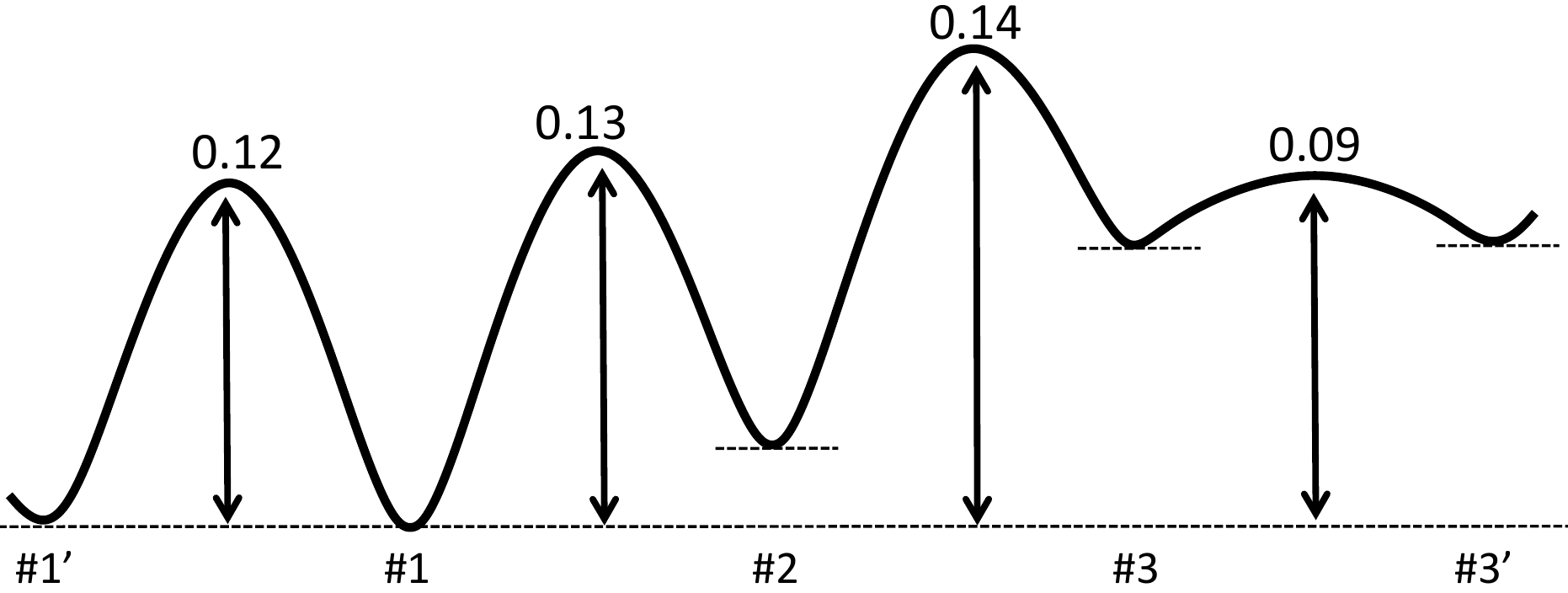}
          \label{fig:migbarriers}
}
       \subfigure{
        \includegraphics[height=0.8\myhgt]{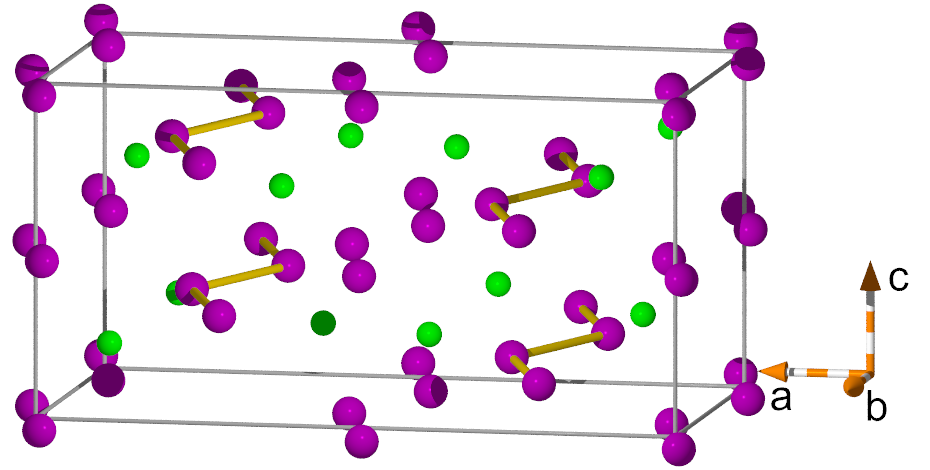}
        \label{fig:onetoone}
}
      \subfigure{
        \includegraphics[height=0.8\myhgt]{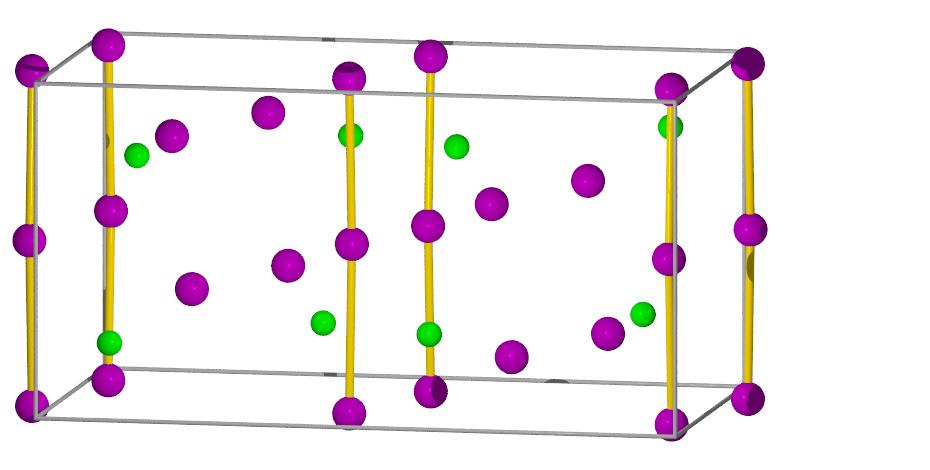}
        \label{fig:threetothree}
}
  \caption{
    Adiabatic potential energy landscape for polaron migration.
    (a) Matrix of migration barriers (in eV) separating different dimer configurations.
    Empty cells in the matrix indicate dimer combinations that do not share an iodine ion and hence are not directly connected.
    (b) Energy landscape and (c) \textbf{1}$\rightarrow$\textbf{1'}, (d)\textbf{3}$\rightarrow$\textbf{3'} atomic scale representation of efficient (low energy) polaron migration pathways.
  }
  \label{fig:barriers}
\end{figure*}

There exist 66 symmetrically distinct A--B$\rightarrow$A--C paths that share one I ion and thus {\em directly} connect two dimer configurations [\fig{fig:table}]. Four paths bridge between symmetry-equivalent dimers, such as \textbf{1}$\rightarrow$\textbf{1'} and \textbf{3}$\rightarrow$\textbf{3'} [\fig{fig:barriers}(c-d)]. Most of the transition barriers fall between 0.1 and 0.4\,eV with a few paths exhibiting very high barriers of more than 0.6\,eV. 

The three most stable dimers (\textbf{1}--\textbf{3}) are connected by network with particular low barriers [\fig{fig:barriers}(c-d)].
The most stable configurations (\textbf{1}) are directly connected via a zig-zag path along the $c$-axis [\Fig{fig:onetoone}] with a barrier height of 0.12\,eV. While already this result suggests that polaron diffusion in SrI$_2$ is much faster than in e.g., sodium iodide where the lowest migration barrier is 0.19\,eV \cite{SadErhAbe15}, the availability of additional pathways further separates these materials.
Specifically, in thermal equilibrium at 300\,K, the population of type \textbf{3} polarons as given by the Boltzmann factor is approximately 5\%. Polarons of this type are accessible from configuration \textbf{1} via \textbf{2} with an effective barrier of 0.14\,eV [\fig{fig:migbarriers}]. They are furthermore connected with each other via a path parallel to the $a$-axis with a barrier of only 0.03\,eV [\fig{fig:threetothree}], which enables practically barrier free diffusion at room temperature. Interestingly, the barrier for migration via the \textbf{3}$\rightarrow$\textbf{3'} network is smaller (0.03,\eV) than for the ``recombination'' reaction \textbf{3}$\rightarrow$\textbf{2} (0.08\,eV). These APES features imply that already at room temperature polarons in SrI$_2$ can diffuse extremely fast with one-dimensional characteristics. This situation is reminiscent of the kick-out mechanism leading to fast diffusion of gold in silicon. Here the solubility of Au on an interstitial site is lower than on a regular lattice site but with the reversed relation between their mobilities.\cite{GosFraSee80} 

While the foregoing considerations apply {\em under equilibrium conditions}, in the following section, we will demonstrate that {\em under irradiation conditions} one can expect an even larger polaron mobility due to an effective inversion of the equilibrium population.

\subsection{Polaron self-trapping and equilibration}

In this section, we address ({\em i}) the time-scale of polaron self-trapping, ({\em ii}) the nascent distribution of polaron configurations, and ({\em iii}) its evolution toward thermodynamical equilibrium based on adiabatic molecular dynamics (MD) simulations.

\begin{figure}
  \includegraphics[scale=0.62]{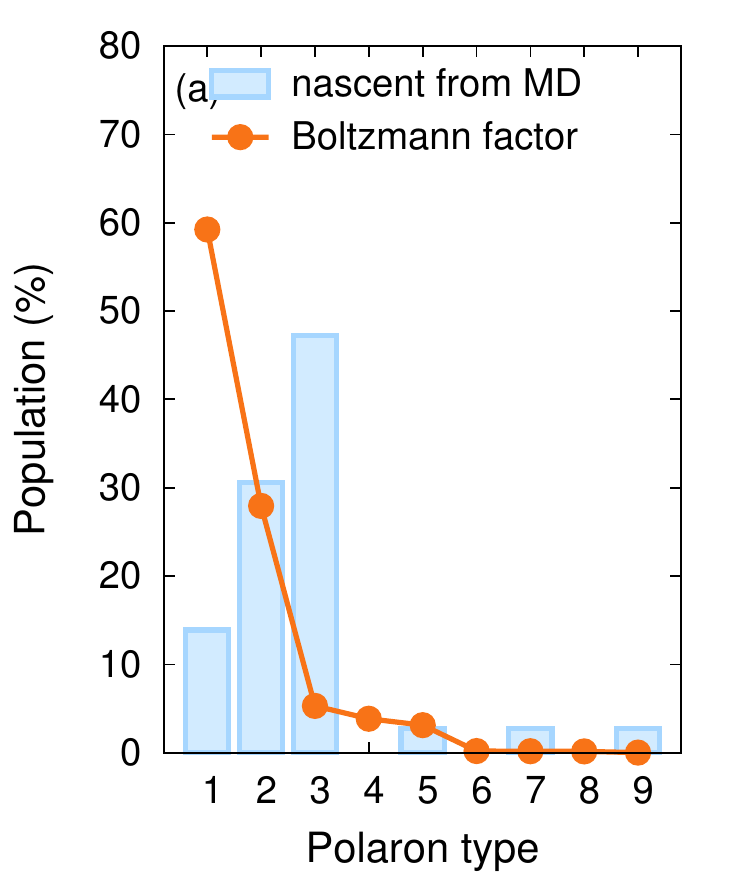}%
  \includegraphics[scale=0.62,clip=10 0 0 0]{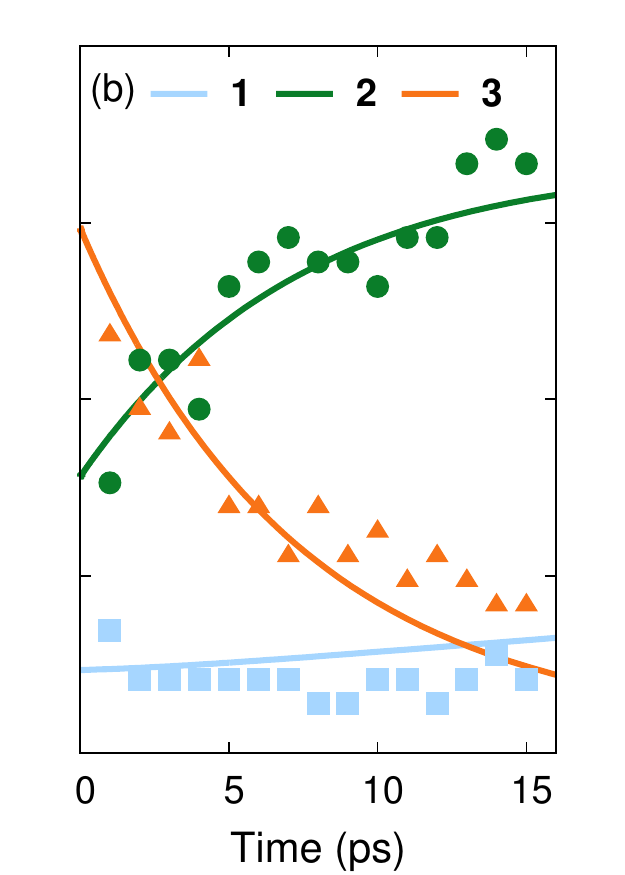}
  \caption{
    (a) Population of polaron configurations at 300\,K as given by the Boltzmann factor $\exp(-\Delta G_f/k_B T)$ with $\Delta G_f \approx \Delta E_f$ as well as the nascent population obtained by analyzing MD trajectories.
    (b) Evolution of the polaron population for three dominant polaron types with time at 300\,K as obtained by averaging of MD trajectories. The lines are guides to the eye.
  }
  \label{fig:polaron-distribution}
\end{figure}

\begin{figure}
    \centering
    \includegraphics[width=0.58\columnwidth]{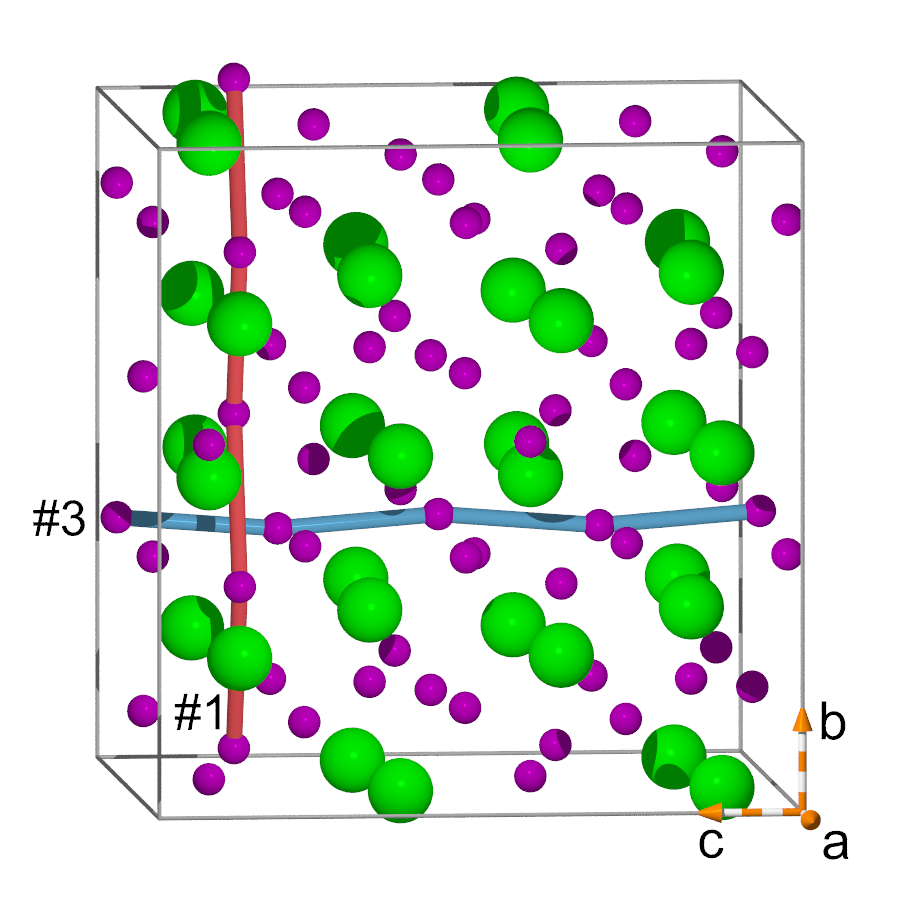}
    \includegraphics[width=0.92\columnwidth]{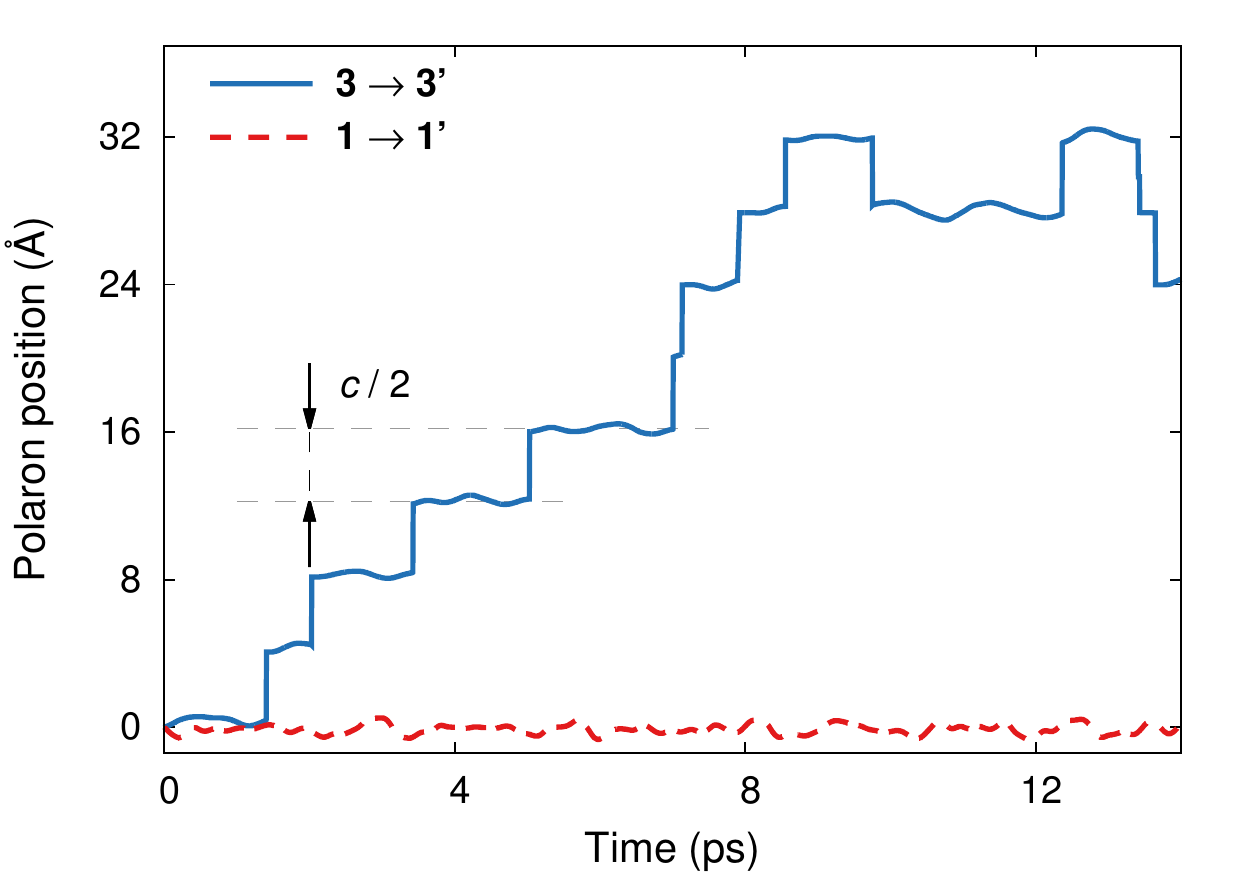}
\caption{ (a) Perspective view of SrI$_2$, illustrating \textbf{1}$\rightarrow$\textbf{1'} (red) and \textbf{3}$\rightarrow$\textbf{3'} jumps (blue). (b) Polaron dynamics for type \textbf{1} and \textbf{3} polarons at room temperature. The metastable \textbf{3} configuration rapidly moves along the $b$-axis at a rate of almost one jump per picosecond. By contrast, no jumps are observed on the time scale of the MD simulations (about 15\,ps) for the most stable configuration \textbf{1}, which preferentially migrates parallel to the $b$-axis.}
    \label{fig:polaron-dynamics}
\end{figure}

In the MD simulations, self-trapping can be observed via the simultaneous ejection of an antibonding $5p\sigma^*$ hole state into the gap and a shortening of the bond length of an iodine pair to about 3.3--3.4\,\AA{} (compare Fig.~1 in the Supplementary Material). It is apparent that self-trapping occurs almost immediately, on the sub-picosecond time scale (see Fig.~1 in the Supplementary Material). 
The fluctuations in the nearest-neighbor iodine-dimer distances are thus sufficiently large to readily relax into a polaronic configuration in the presence of a hole charge and the self-trapping process is effectively barrier free. 

By averaging over many MD trajectories it is possible to determine the distribution of ``as-born'' (nascent) carriers over different polaron types [\fig{fig:polaron-distribution}(a)]. Compared to the equilibrium distribution, which for simplicity can be approximated by assuming $\Delta G_f \approx \Delta E_f$, it is striking that the probability for polarons of types \textbf{1} (59\%\ in equilibrium \textit{vs.} 14\% nascent) and \textbf{3} (5\%\ \textit{vs.} 47\%) is practically inverted. At room temperature polarons of type \textbf{3} are gradually converted to type \textbf{2} over the time scale of 5 to 15\,ps, while the number of type \textbf{1} polarons remains low [\fig{fig:polaron-distribution}(b)]. This is compatible with the lack of a direct pathway between configurations \textbf{3} and \textbf{1} [\fig{fig:migbarriers}]. 

The population inversion has important consequences for polaron migration. The barrier for \textbf{3}$\rightarrow$\textbf{3'} jumps is only 30\,meV [\fig{fig:migbarriers}] and thus significantly smaller than for \textbf{1}$\rightarrow$\textbf{1'} events (120\,meV). (Note that there is no \emph{direct} pathway between polarons of type \textbf{2}). Under equilibrium conditions type \textbf{1} polarons outnumber polarons of type \textbf{3}, which outweighs the lower barriers available to the latter. The population inversion in the nascent distribution, however, implies that migration via the \textbf{3}$\rightarrow$\textbf{3'} pathway is strongly enhanced. This picture is confirmed by explicit MD simulations, which show extremely rapid and practically athermal diffusion of type \textbf{1} polarons [\fig{fig:polaron-dynamics}] with tens of hopping events occurring over just 15\,ps at room temperature, whereas the polarons of type \textbf{1} remain stationary under the same conditions.

\section*{Discussion}
\label{sec:conclusion}

Based on the results presented in the previous section, we will in the following argue that both the inversion of the equilibrium distribution and the rapid migration of type \textbf{3} polarons are pivotal for the high light-yield proportionality of SrI$_2$. To this end,  we first briefly review the scintillation process. 

The incident radiation initially produces fast electrons, either directly by fundamental light-matter interactions or indirectly via electron-electron scattering, that traverse the crystal and deposit their energy by exciting electron-hole pairs and plasmons.\cite{Rod97, Mos02, Pay15} This process occurs on the femtosecond timescale and gives rise to a distribution of low-energy carriers in a nanometer-sized region surrounding the electrons tracks. The excitation density increases with decreasing kinetic energy of the fast electron. \cite{MurMey61, DorHaaEij95}
Nevertheless, at this point the total number of low-energy electron-hole pairs is approximately proportional to the initial photon energy. In an ideal scintillator without losses, all pairs would transfer their energy to the activator ions, resulting in a photon count that is proportional to the incident energy. In real materials, the low-energy excitations and carriers are, however, subject to annihilation via non-radiative quenching mechanisms such as Auger recombination of free carriers or exciton-exciton annihilation.\cite{Vas08} As the rates of these processes exhibit a non-linear dependence on electron-hole or exciton density,\cite{Vas08} the non-uniformity of the excitation density along the track will give rise to a total light yield that is no longer proportional to the energy of the incident photon energy. It is also apparent that the non-radiative recombination rates are maximal during the initial stage, when the excitation density is largest. In particular, it has been conjectured \cite{LiGriWil11} that the first 10 picoseconds are the most decisive for the proportionality of the light-yield response. Thus, reduction of the excitation density would naturally lead to reduced quenching and increased proportionality. This can be accomplished by means of temporary carrier capture on shallow traps as discussed by Williams \textit{et al.} \cite{LiGriWil11,WilGriLi15} in the context of Tl-doped CsI. The same effect was obtained by co-doping Ce and Sr in lanthanum bromide \cite{AleHaaKho13, AbeSadSch14, ErhSadSch15}.

Extending this idea to deep traps, such as self-trapped polarons, has so far not been considered useful. In common halides materials, such as NaI and CsI, self-trapped (hole) polarons feature both large binding and migration energies, a coupling that is expected based on the structure of the potential energy surface. Polarons therefore exhibit both long detrapping times and slow migration. However, as demonstrated in this work, self-trapped polarons in SrI$_2$
\begin{enumerate}
\item[(\emph{i})]
    are characterized by trapping times below 1\,ps,
\item[(\emph{ii})]
    have large binding energies comparable to other halides,
\item[(\emph{iii})]
    are predominantly ``born'' as type \textbf{3} polarons, and
\item[(\emph{iii})]
    in the latter configuration exhibit essentially barrier-free migration at room temperature.
\end{enumerate}
The existence of multiple low energy configurations (which is possible due to the low symmetry of the system) in combination with a comparably soft energy landscape (compare the phonon dispersion) gives rise to a nascent polaron population that strongly favors highly mobile polaron configurations (\textbf{3}). We speculate that these polarons can quickly migrate away from the initial track region while remaining localized (limiting Auger recombination), causing the track to broaden more quickly. This in turn reduces the density of electron-hole pairs during the crucial first picoseconds, and thus lowers the rates of the non-linear non-radiative recombination processes that constitute the root cause of the non-linear response. This mechanism provides a rationale for the superior proportionality of SrI$_2$ compared to materials such as NaI and CsI.

To contrast the polaron dynamics in SrI$_2$ and NaI, a material with rather poor energy resolution, we have also performed molecular-dynamics simulations of polaron formation and transport in this material at 300\,K using 216 atom supercells. These calculations confirm that even in NaI, self-trapping occurs at sub-picosecond time scale at room temperature. However, we never observe a hopping event in any of our NaI simulations, regardless of initial conditions. It is thus evident that polaron transport is much slower in NaI as compared with SrI$_2$. 

The present work provides the basis for future large-scale simulations based on e.g., rate-equation models \cite{Vas08, WilGriLi15} or kinetic Monte-Carlo methods \cite{KerRosCan09}, which enable quantitative simulations of light yield based on a set of materials specific input parameters.
In this context, we close by commenting on the importance of using suitable methods to obtain the relevant rates: Assuming identical pre-factors, one would expect the conversion between \textbf{2} and \textbf{1} ($\Delta E_m = 0.10\,\eV$) to take approximately 4 to 5 times longer ($\exp[\Delta E_m^{2\rightarrow 1}/\Delta E_m^{3\rightarrow2}]$) than from \textbf{3} to \textbf{2} ($\Delta E_m = 0.08\,\eV$). To match the MD data [\fig{fig:polaron-distribution}(b)] one, however, requires a ratio that is about 20--40 times larger. This deviation emphasizes the role of vibrational degrees of freedom, which affect the attempt frequencies and even more so the formation \emph{free} energies. It thereby highlights the importance of direct MD simulations, as opposed to e.g., \emph{solely} harmonic transition state theory, for understanding polaron dynamics.

\section*{Methods}
The failure of DFT to stabilize polarons in wide gap insulators is related to the self-interaction present in common semi-local exchange-correlation (XC) functionals \cite{Gavartin2003PRB035108}, which manifests itself in the convexity of the total energy as a function of the fractional hole charge \cite{LanZun09, Erhart2014PRB035204}. The recently developed pSIC method \cite{SadErhAbe15} builds on this insight and provides a parameter-free approach for studying polaronic properties that is both accurate and computationally efficient. The pSIC total energy functional for a system with an added electron/hole is given by
\begin{align}
E_{\text{pSIC}}^{\pm}[\vec{R}] = E_{\text{DFT}}[0;\vec{R}] \pm \mu_{\text{DFT}}^{\pm}[\vec{R}],
\label{eq:psic}
\end{align}
where $E_{\text{DFT}}[0;\vec{R}]$ signifies the total energy of the ionic configuration $\vec{R}$, in the reference electronic charge state, which for applications in this report is chosen as the {\em neutral} charge state. Furthermore, $\mu_{\text{DFT}}^{\pm}$ is the electron chemical potential, {\it i.e.} the right/left derivative of $E_{\text {DFT}}$ with respect to electron number at the reference electronic configuration. The main approximation in \eq{eq:psic} stems from the omission of the largely $\vec{R}$-independent many-body scissors corrections to the DFT band energies at the valence band maximum and the conduction band minimum. The pSIC predicted migration barrier for hole polarons in NaI was in good agreement with experiment \cite{SadErhAbe15}, encouraging the present comprehensive study of SrI$_2$.

SrI$_2$ with its space-group Pbca (No. 61 in the International Tables of Crys\-tal\-lo\-graphy \cite{ITC}) contains 24 atoms  in the primitive unit cell with one Sr and two I sub-lattices, each occupying an 8c Wyckoff site with three degrees of freedom [Fig.~\ref{fig:structure}(a)]. We used experimental values for the cell metric: $a = 15.22\,\text{\AA}$, $b = 8.22\,\text{\AA}$, and $c = 7.90\,\text{\AA}$ \cite{Barnighausen1969ZK430}.  All internal coordinates are fully relaxed in our DFT calculations  using the PBE parametrization of the generalized gradient approximation (GGA), projector augmented-wave (PAW) potentials \cite{Blochl1994PRB17953, *Kresse1999PRB1758}, a cutoff energy of 229\,eV, and no symmetry constraints as implemented in the Vienna ab-initio simulation package \cite{Kresse1999PRB1758}. Monoclinic $(a,0,0) \times (0, b, -c) \times (0, b, 2c)$ supercells containing 72 atoms were constructed to study polaron formation and migration together with a $2 \times 3 \times 2$ $\vec{k}$-point sampling. For the larger cells used in the size-convergence study (up to 960 atoms) only the $\Gamma$-point was included (see Supporting Information). Atomic positions were optimized until the residual forces were less than 0.01\,meV/\AA.

In the pSIC method, polaron formation energies are calculated according to
\begin{align}
\Delta E_f[\vec{R}_{\text{pol}}] = E^{\pm}_\text{pSIC}[\vec{R}_{\text{ideal}}]  - E^{\pm}_\text{pSIC}[\vec{R}_{\text{pol}}] ,
\end{align}
where $\vec{R}_{\text{ideal}}$ and $\vec{R}_{\text{pol}}$ denote the ideal and a polaron configuration, respectively. As this method relies on electron chemical potentials computed in the neutral reference state, it is not subject to image charge corrections. The pSIC results were benchmarked against PBE0 hybrid calculations, from which $V_K$-center formation energies are obtained as energy differences between {\em charged} and {\em neutral} supercells \cite{Lany2008PRB235104, Erhart2014PRB035204} according to
\begin{align}
\Delta E_f &= E_{\text{PBE0}}[0, \vec{R}_{\text{ideal}}] - E_{\text{PBE0}}[\pm 1, \vec{R}_{\text{pol}}] \pm E_{\text{VBM}}.
\end{align}

Image charge corrections are applied to hybrid calculations with the modified Makov-Payne correction of Lany and Zunger \cite{Makov1995PRB4014, *Lany2008PRB235104}. To this end, we computed both the ionic and electronic contributions to the dielectric constant. In the latter case, we approximately included local field effects for the hybrid functionals by adding the difference between DFT dielectric constants with and without local field effects. In this fashion, we obtained the diagonal dielectric tensor $\epsilon = \text{diag}(6.42, 6.13, 7.09)$ resulting in a monopole-monopole correction of 0.207\,eV.

To study the polaron migration between dimers A--B$\rightarrow$A--C, a systematic exploration of symmetrically distinct iodine triplets A--B--C was performed where $\I_\A$--$\I_\B$ and $\I_\A$--$\I_\C$ are stable dimers. Here, two pairs (triplets) are considered equivalent if there exists a space group operation that maps A--B to A'--B' (A--B--C to A'--B'--C'). The search for polaron migration paths was performed using the Compressive Sensing Lattice Dynamics (CSLD) code \cite{Zhou2014PRL185501}. The corresponding minimum energy path (MEP) for each A--B$\rightarrow$A--C combination was optimized with the nudged elastic band  (NEB) method \cite{Henkelman2000JCP9901,Henkelman2000JCP9978} using three intermediate images.


MD simulations were performed at 300\,K using $(a,0,0)\times(0,2b,0)\times(0,0,2c)$ (96 atom) supercells and an integration time step of 4\,fs. The temperature of the system was controlled via a Nos\'e-Hoover thermostat with the Nos\'e mass chosen such that the temperature fluctuates with a period of about 0.1\,ps. We generated 36 separate MD trajectories of the system with a hole charge, the initial ionic coordinates and velocities for which were taken from an equilibration run carried out for the neutral system. The pSIC functional was used to calculate the ionic forces in the presence of the charge excitation. The self-trapping process deposits a large amount of energy in a small volume element equivalent to the polaron binding energy. For the polarons in SrI$_2$ this amounts to about 0.36\,eV in the 96-atom supercell that has been used for MD studies. This heat has to be dissipated before equilibrium can be attained. The Nos\'e-Hoover thermostat used here equilibrates the temperature of the system within 0.1\,ps, which is faster than in the real system. As a result, our simulations tend to underestimate the extent of the non-equilibrium character observed in the MD studies reported in this section. MD duration was limited to $\lesssim$ 20 ps to focus on polaron dynamics most relevant to the scintillator response.

\section*{ACKNOWLEDGEMENTS}
Fruitful discussions with S. A. Payne and N. J. Cherepy are acknowledged. This work was performed under the auspices of the U.S. Department of Energy by Lawrence Livermore National Laboratory under Contract No. DE- AC52-07NA27344 with the support from the National Nuclear Security Administration Office of Nonproliferation Research and Development (NA-22). One of us (PE) acknowledges funding from the Knut and Alice Wallenberg foundation in the form of a fellowship.

\section*{CONTRIBUTIONS}
All authors contributed extensively to the work presented in this paper.

\section*{COMPETING INTERESTS}
The authors declare no conflict of interest.


\begin{thebibliography}{10}
	\expandafter\ifx\csname url\endcsname\relax
	\def\url#1{\texttt{#1}}\fi
	\expandafter\ifx\csname urlprefix\endcsname\relax\def\urlprefix{URL }\fi
	\providecommand{\bibinfo}[2]{#2}
	\providecommand{\eprint}[2][]{\url{#2}}
	
	\bibitem{WilSon90}
	\bibinfo{author}{Williams, R.~T.} \& \bibinfo{author}{Song, K.~S.}
	\newblock \bibinfo{title}{{The self-trapped exciton}}.
	\newblock \emph{\bibinfo{journal}{Journal of Physics and Chemistry of Solids}}
	\textbf{\bibinfo{volume}{51}}, \bibinfo{pages}{679--716}
	(\bibinfo{year}{1990}).
	
	\bibitem{MaxZhoCed06}
	\bibinfo{author}{Maxisch, T.}, \bibinfo{author}{Zhou, F.} \&
	\bibinfo{author}{Ceder, G.}
	\newblock \bibinfo{title}{{Ab initio study of the migration of small polarons
			in olivine {LixFePO4} and their association with lithium ions and
			vacancies}}.
	\newblock \emph{\bibinfo{journal}{Phys. Rev. B}} \textbf{\bibinfo{volume}{73}},
	\bibinfo{pages}{104301} (\bibinfo{year}{2006}).
	
	\bibitem{DeaAraFra12}
	\bibinfo{author}{De\'ak, P.}, \bibinfo{author}{Aradi, B.} \&
	\bibinfo{author}{Frauenheim, T.}
	\newblock \bibinfo{title}{Quantitative theory of the oxygen vacancy and carrier
		self-trapping in bulk {TiO$_{2}$}}.
	\newblock \emph{\bibinfo{journal}{Phys. Rev. B}} \textbf{\bibinfo{volume}{86}},
	\bibinfo{pages}{195206} (\bibinfo{year}{2012}).
	
	\bibitem{Erhart2014PRB035204}
	\bibinfo{author}{Erhart, P.}, \bibinfo{author}{Klein, A.},
	\bibinfo{author}{{\AA}berg, D.} \& \bibinfo{author}{Sadigh, B.}
	\newblock \bibinfo{title}{{Efficacy of the DFT+U formalism for modeling hole
			polarons in perovskite oxides}}.
	\newblock \emph{\bibinfo{journal}{Phys. Rev. B}} \textbf{\bibinfo{volume}{90}},
	\bibinfo{pages}{035204} (\bibinfo{year}{2014}).
	
	\bibitem{Kanzig1955PR1890}
	\bibinfo{author}{K{\"a}nzig, W.}
	\newblock \bibinfo{title}{{Electron Spin Resonance of V1-Centers}}.
	\newblock \emph{\bibinfo{journal}{Phys. Rev.}} \textbf{\bibinfo{volume}{99}},
	\bibinfo{pages}{1890} (\bibinfo{year}{1955}).
	
	\bibitem{Gavartin2003PRB035108}
	\bibinfo{author}{Gavartin, J.~L.}, \bibinfo{author}{Sushko, P.~V.} \&
	\bibinfo{author}{Shluger, A.~L.}
	\newblock \bibinfo{title}{{Modeling charge self-trapping in wide-gap
			dielectrics: Localization problem in local density functionals}}.
	\newblock \emph{\bibinfo{journal}{Phys. Rev. B}} \textbf{\bibinfo{volume}{67}},
	\bibinfo{pages}{035108} (\bibinfo{year}{2003}).
	
	\bibitem{SadErhAbe15}
	\bibinfo{author}{Sadigh, B.}, \bibinfo{author}{Erhart, P.} \&
	\bibinfo{author}{{\AA}berg, D.}
	\newblock \bibinfo{title}{{Variational polaron self-interaction-corrected
			total-energy functional for charge excitations in insulators}}.
	\newblock \emph{\bibinfo{journal}{Phys. Rev. B}} \textbf{\bibinfo{volume}{92}},
	\bibinfo{pages}{075202} (\bibinfo{year}{2015}).
	\newblock \bibinfo{note}{Erratum, {\it ibid.} \textbf{92}, 199905(E) (2015)}.
	
	\bibitem{Anisimov1991PRB943}
	\bibinfo{author}{Anisimov, V.~I.}, \bibinfo{author}{Zaanen, J.} \&
	\bibinfo{author}{Andersen, O.~K.}
	\newblock \bibinfo{title}{{Band Theory and Mott Insulators - Hubbard-U Instead
			of Stoner-I}}.
	\newblock \emph{\bibinfo{journal}{Phys. Rev. B}} \textbf{\bibinfo{volume}{44}},
	\bibinfo{pages}{943--954} (\bibinfo{year}{1991}).
	
	\bibitem{LanZun09}
	\bibinfo{author}{Lany, S.} \& \bibinfo{author}{Zunger, A.}
	\newblock \bibinfo{title}{{Polaronic hole localization and multiple hole
			binding of acceptors in oxide wide-gap semiconductors}}.
	\newblock \emph{\bibinfo{journal}{Phys. Rev. B}} \textbf{\bibinfo{volume}{80}},
	\bibinfo{pages}{085202} (\bibinfo{year}{2009}).
	
	\bibitem{BisDu12}
	\bibinfo{author}{Biswas, K.} \& \bibinfo{author}{Du, M.-H.}
	\newblock \bibinfo{title}{{Energy transport and scintillation of cerium-doped
			elpasolite Cs$_2$LiYCl$_6$: Hybrid density functional calculations}}.
	\newblock \emph{\bibinfo{journal}{Phys. Rev. B}} \textbf{\bibinfo{volume}{86}},
	\bibinfo{pages}{014102} (\bibinfo{year}{2012}).
	
	\bibitem{CheHulDro08}
	\bibinfo{author}{Cherepy, N.~J.} \emph{et~al.}
	\newblock \bibinfo{title}{{Strontium and barium iodide high light yield
			scintillators}}.
	\newblock \emph{\bibinfo{journal}{Appl. Phys. Lett.}}
	\textbf{\bibinfo{volume}{92}} (\bibinfo{year}{2008}).
	
	\bibitem{WilLoeGlo08}
	\bibinfo{author}{Wilson, C.~M.} \emph{et~al.}
	\newblock \bibinfo{title}{{Strontium iodide scintillators for high energy
			resolution gamma ray spectroscopy}}.
	\newblock \emph{\bibinfo{journal}{Proc. SPIE}} \textbf{\bibinfo{volume}{7079}}
	(\bibinfo{year}{2008}).
	
	\bibitem{ChePayAsz09}
	\bibinfo{author}{Cherepy, N.~J.} \emph{et~al.}
	\newblock \bibinfo{title}{{Scintillators With Potential to Supersede Lanthanum
			Bromide}}.
	\newblock \emph{\bibinfo{journal}{IEEE Trans. Nucl. Sci.}}
	\textbf{\bibinfo{volume}{56}}, \bibinfo{pages}{873--880}
	(\bibinfo{year}{2009}).
	
	\bibitem{AleKhoHaa12}
	\bibinfo{author}{Alekhin, M.}, \bibinfo{author}{Khodyuk, I.},
	\bibinfo{author}{de~Haas, J.} \& \bibinfo{author}{Dorenbos, P.}
	\newblock \bibinfo{title}{{Nonproportional Response and Energy Resolution of
			Pure SrI$_{2}$ and SrI$ _{2}$:5\%Eu Scintillators}}.
	\newblock \emph{\bibinfo{journal}{IEEE Trans. Nucl. Sci.}}
	\textbf{\bibinfo{volume}{59}}, \bibinfo{pages}{665--670}
	(\bibinfo{year}{2012}).
	
	\bibitem{ErhSchSad14}
	\bibinfo{author}{Erhart, P.}, \bibinfo{author}{Schleife, A.},
	\bibinfo{author}{Sadigh, B.} \& \bibinfo{author}{{\AA}berg, D.}
	\newblock \bibinfo{title}{{Quasiparticle spectra, absorption spectra, and
			excitonic properties of NaI and SrI$_2$ from many-body perturbation theory}}.
	\newblock \emph{\bibinfo{journal}{Phys. Rev. B}} \textbf{\bibinfo{volume}{89}},
	\bibinfo{pages}{075132} (\bibinfo{year}{2014}).
	
	\bibitem{LoeDorEij01}
	\bibinfo{author}{van Loef, E. V.~D.}, \bibinfo{author}{Dorenbos, P.},
	\bibinfo{author}{van Eijk, C. W.~E.}, \bibinfo{author}{Kr\"amer, K.} \&
	\bibinfo{author}{G\"udel, H.~U.}
	\newblock \bibinfo{title}{{High-energy-resolution scintillator: Ce$^{3+}$
			activated LaBr$_3$}}.
	\newblock \emph{\bibinfo{journal}{Appl. Phys. Lett.}}
	\textbf{\bibinfo{volume}{79}}, \bibinfo{pages}{1573} (\bibinfo{year}{2001}).
	
	\bibitem{AbeSadErh12}
	\bibinfo{author}{{\AA}berg, D.}, \bibinfo{author}{Sadigh, B.} \&
	\bibinfo{author}{Erhart, P.}
	\newblock \bibinfo{title}{{Electronic structure of {LaBr$_3$} from
			quasiparticle self-consistent {\it GW} calculations}}.
	\newblock \emph{\bibinfo{journal}{Phys. Rev. B}} \textbf{\bibinfo{volume}{85}},
	\bibinfo{pages}{125134} (\bibinfo{year}{2012}).
	
	\bibitem{PawSpa97}
	\bibinfo{author}{Pawlik, T.} \& \bibinfo{author}{Spaeth, J.-M.}
	\newblock \bibinfo{title}{{Electron and hole centres in the X-irradiated
			elpasolite crystal studied by means of electron paramagnetic resonance and
			electron nuclear double resonance}}.
	\newblock \emph{\bibinfo{journal}{J. Phys. Cond. Matter}}
	\textbf{\bibinfo{volume}{9}}, \bibinfo{pages}{8737} (\bibinfo{year}{1997}).
	
	\bibitem{BesDorEij04}
	\bibinfo{author}{Bessi\`ere, A.} \emph{et~al.}
	\newblock \bibinfo{title}{{Spectroscopy and anomalous emission of {Ce} doped
			elpasolite Cs$_2$LiYCl$_6$}}.
	\newblock \emph{\bibinfo{journal}{J. Phys. Cond. Matter}}
	\textbf{\bibinfo{volume}{16}}, \bibinfo{pages}{1887} (\bibinfo{year}{2004}).
	
	\bibitem{Perdew1996PRL3865}
	\bibinfo{author}{Perdew, J.~P.}, \bibinfo{author}{Burke, K.} \&
	\bibinfo{author}{Ernzerhof, M.}
	\newblock \bibinfo{title}{{Generalized gradient approximation made simple}}.
	\newblock \emph{\bibinfo{journal}{Phys. Rev. Lett.}}
	\textbf{\bibinfo{volume}{77}}, \bibinfo{pages}{3865--3868}
	(\bibinfo{year}{1996}).
	
	\bibitem{Adamo1999JCP6158}
	\bibinfo{author}{Adamo, C.} \& \bibinfo{author}{Barone, V.}
	\newblock \bibinfo{title}{{Toward reliable density functional methods without
			adjustable parameters: The PBE0 model}}.
	\newblock \emph{\bibinfo{journal}{J. Chem. Phys.}}
	\textbf{\bibinfo{volume}{110}}, \bibinfo{pages}{6158} (\bibinfo{year}{1999}).
	
	\bibitem{GosFraSee80}
	\bibinfo{author}{G\"osele, U.}, \bibinfo{author}{Frank, W.} \&
	\bibinfo{author}{Seeger, A.}
	\newblock \bibinfo{title}{{Mechanism and kinetics of the diffusion of gold in
			silicon}}.
	\newblock \emph{\bibinfo{journal}{Applied physics}}
	\textbf{\bibinfo{volume}{23}}, \bibinfo{pages}{361--368}
	(\bibinfo{year}{1980}).
	
	\bibitem{Rod97}
	\bibinfo{author}{Rodnyi, P.~A.}
	\newblock \emph{\bibinfo{title}{{Physical processes in inorganic
				scintillators}}} (\bibinfo{publisher}{{CRC} Press}, \bibinfo{address}{Boca
		Raton}, \bibinfo{year}{1997}).
	
	\bibitem{Mos02}
	\bibinfo{author}{Moses, W.~W.}
	\newblock \bibinfo{title}{{Current trends in scintillator detectors and
			materials}}.
	\newblock \emph{\bibinfo{journal}{Nucl. Instrum. Meth. A}}
	\textbf{\bibinfo{volume}{487}}, \bibinfo{pages}{123 -- 128}
	(\bibinfo{year}{2002}).
	\newblock \bibinfo{note}{3rd International Workshop on Radiation Imaging
		Detectors}.
	
	\bibitem{Pay15}
	\bibinfo{author}{Payne, S.~A.}
	\newblock \bibinfo{title}{{Nonproportionality of Scintillator Detectors. IV.
			Resolution Contribution from Delta-Rays}}.
	\newblock \emph{\bibinfo{journal}{IEEE Trans. Nucl. Sci.}}
	\textbf{\bibinfo{volume}{62}}, \bibinfo{pages}{372--380}
	(\bibinfo{year}{2015}).
	
	\bibitem{MurMey61}
	\bibinfo{author}{Murray, R.~B.} \& \bibinfo{author}{Meyer, A.}
	\newblock \bibinfo{title}{{Scintillation Response of Activated Inorganic
			Crystals to Various Charged Particles}}.
	\newblock \emph{\bibinfo{journal}{Phys. Rev.}} \textbf{\bibinfo{volume}{122}},
	\bibinfo{pages}{815--826} (\bibinfo{year}{1961}).
	
	\bibitem{DorHaaEij95}
	\bibinfo{author}{Dorenbos, P.}, \bibinfo{author}{de~Haas, J. T.~M.} \&
	\bibinfo{author}{van Eijk, C. W.~E.}
	\newblock \bibinfo{title}{{Non-proportionality in the scintillation response
			and the energy resolution obtainable with scintillation crystals}}.
	\newblock \emph{\bibinfo{journal}{IEEE Trans. Nucl. Sci.}}
	\textbf{\bibinfo{volume}{42}}, \bibinfo{pages}{2190--2202}
	(\bibinfo{year}{1995}).
	
	\bibitem{Vas08}
	\bibinfo{author}{Vasil'ev, A.~V.}
	\newblock \bibinfo{title}{{From Luminescence Non-Linearity to Scintillation
			Non-Proportionality}}.
	\newblock \emph{\bibinfo{journal}{IEEE Trans. Nucl. Sci.}}
	\textbf{\bibinfo{volume}{55}}, \bibinfo{pages}{1054} (\bibinfo{year}{2008}).
	
	\bibitem{LiGriWil11}
	\bibinfo{author}{Li, Q.}, \bibinfo{author}{Grim, J.~Q.},
	\bibinfo{author}{Williams, R.~T.}, \bibinfo{author}{Bizarri, G.~A.} \&
	\bibinfo{author}{Moses, W.~W.}
	\newblock \bibinfo{title}{{A transport-based model of material trends in
			nonproportionality of scintillators}}.
	\newblock \emph{\bibinfo{journal}{J. Appl. Phys.}}
	\textbf{\bibinfo{volume}{109}} (\bibinfo{year}{2011}).
	
	\bibitem{WilGriLi15}
	\bibinfo{author}{Williams, R.~T.} \emph{et~al.}
	\newblock \emph{\bibinfo{title}{{Excitonic and Photonic Processes in
				Materials}}}, chap. \bibinfo{chapter}{Scintillation Detectors of Radiation:
		Excitations at High Densities and Strong Gradients},
	\bibinfo{pages}{299--358} (\bibinfo{publisher}{Springer Singapore},
	\bibinfo{address}{Singapore}, \bibinfo{year}{2015}).
	
	\bibitem{AleHaaKho13}
	\bibinfo{author}{Alekhin, M.~S.} \emph{et~al.}
	\newblock \bibinfo{title}{{Improvement of $\gamma$-ray energy resolution of
			LaBr$_3$:Ce$^{3+}$ scintillation detectors by Sr$^{2+}$ and Ca$^{2+}$
			co-doping}}.
	\newblock \emph{\bibinfo{journal}{Appl. Phys. Lett.}}
	\textbf{\bibinfo{volume}{102}}, \bibinfo{pages}{161915}
	(\bibinfo{year}{2013}).
	
	\bibitem{AbeSadSch14}
	\bibinfo{author}{\AA{}berg, D.}, \bibinfo{author}{Sadigh, B.},
	\bibinfo{author}{Schleife, A.} \& \bibinfo{author}{Erhart, P.}
	\newblock \bibinfo{title}{{Origin of resolution enhancement by co-doping of
			scintillators: Insight from electronic structure calculations}}.
	\newblock \emph{\bibinfo{journal}{Applied Physics Letters}}
	\textbf{\bibinfo{volume}{104}} (\bibinfo{year}{2014}).
	
	\bibitem{ErhSadSch15}
	\bibinfo{author}{Erhart, P.}, \bibinfo{author}{Sadigh, B.},
	\bibinfo{author}{Schleife, A.} \& \bibinfo{author}{\AA{}berg, D.}
	\newblock \bibinfo{title}{{First-principles study of codoping in lanthanum
			bromide}}.
	\newblock \emph{\bibinfo{journal}{Phys. Rev. B}} \textbf{\bibinfo{volume}{91}},
	\bibinfo{pages}{165206} (\bibinfo{year}{2015}).
	
	\bibitem{KerRosCan09}
	\bibinfo{author}{Kerisit, S.}, \bibinfo{author}{Rosso, K.~M.},
	\bibinfo{author}{Cannon, B.~D.}, \bibinfo{author}{Gao, F.} \&
	\bibinfo{author}{Xie, Y.}
	\newblock \bibinfo{title}{{Computer simulation of the light yield nonlinearity
			of inorganic scintillators}}.
	\newblock \emph{\bibinfo{journal}{Journal of Applied Physics}}
	\textbf{\bibinfo{volume}{105}}, \bibinfo{pages}{114915}
	(\bibinfo{year}{2009}).
	
	\bibitem{ITC}
	\bibinfo{author}{Hahn, T.}
	\newblock \emph{\bibinfo{title}{{International Tables for Crystallography,
				Space-Group Symmetry}}}.
	\newblock International Tables for Crystallography (\bibinfo{publisher}{Wiley},
	\bibinfo{address}{New York}, \bibinfo{year}{2005}).
	
	\bibitem{Barnighausen1969ZK430}
	\bibinfo{author}{B\"arnighausen, H.}, \bibinfo{author}{Beck, H.},
	\bibinfo{author}{Grueninger, H.~W.}, \bibinfo{author}{Rietschel, E.~T.} \&
	\bibinfo{author}{Schultz, N.}
	\newblock \bibinfo{title}{{New AB$_2$-type structure with septacoordinated
			cation}}.
	\newblock \emph{\bibinfo{journal}{Z. Krist.}} \textbf{\bibinfo{volume}{128}},
	\bibinfo{pages}{430} (\bibinfo{year}{1969}).
	
	\bibitem{Blochl1994PRB17953}
	\bibinfo{author}{Bl\"ochl, P.~E.}
	\newblock \bibinfo{title}{{Projector augmented-wave method}}.
	\newblock \emph{\bibinfo{journal}{Phys. Rev. B}} \textbf{\bibinfo{volume}{50}},
	\bibinfo{pages}{17953--17979} (\bibinfo{year}{1994}).
	
	\bibitem{Kresse1999PRB1758}
	\bibinfo{author}{Kresse, G.} \& \bibinfo{author}{Joubert, D.}
	\newblock \bibinfo{title}{{From ultrasoft pseudopotentials to the projector
			augmented-wave method}}.
	\newblock \emph{\bibinfo{journal}{Phys. Rev. B}} \textbf{\bibinfo{volume}{59}},
	\bibinfo{pages}{1758--1775} (\bibinfo{year}{1999}).
	
	\bibitem{Lany2008PRB235104}
	\bibinfo{author}{Lany, S.} \& \bibinfo{author}{Zunger, A.}
	\newblock \bibinfo{title}{{Assessment of correction methods for the band-gap
			problem and for finite-size effects in supercell defect calculations: Case
			studies for ZnO and GaAs}}.
	\newblock \emph{\bibinfo{journal}{Phys. Rev. B}} \textbf{\bibinfo{volume}{78}},
	\bibinfo{pages}{235104} (\bibinfo{year}{2008}).
	
	\bibitem{Makov1995PRB4014}
	\bibinfo{author}{Makov, G.} \& \bibinfo{author}{Payne, M.~C.}
	\newblock \bibinfo{title}{{Periodic boundary conditions in ab initio
			calculations}}.
	\newblock \emph{\bibinfo{journal}{Phys. Rev. B}} \textbf{\bibinfo{volume}{51}},
	\bibinfo{pages}{4014} (\bibinfo{year}{1995}).
	
	\bibitem{Zhou2014PRL185501}
	\bibinfo{author}{Zhou, F.}, \bibinfo{author}{Nielson, W.},
	\bibinfo{author}{Xia, Y.} \& \bibinfo{author}{Ozolins, V.}
	\newblock \bibinfo{title}{{Lattice Anharmonicity and Thermal Conductivity from
			Compressive Sensingof First-Principles Calculations}}.
	\newblock \emph{\bibinfo{journal}{Phys. Rev. Lett.}}
	\textbf{\bibinfo{volume}{113}}, \bibinfo{pages}{185501}
	(\bibinfo{year}{2014}).
	
	\bibitem{Henkelman2000JCP9901}
	\bibinfo{author}{Henkelman, G.}, \bibinfo{author}{Uberuaga, B.~P.} \&
	\bibinfo{author}{Jonsson, H.}
	\newblock \bibinfo{title}{{A climbing image nudged elastic band method for
			finding saddle points and minimum energy paths}}.
	\newblock \emph{\bibinfo{journal}{J. Chem. Phys.}}
	\textbf{\bibinfo{volume}{113}}, \bibinfo{pages}{9901--9904}
	(\bibinfo{year}{2000}).
	
	\bibitem{Henkelman2000JCP9978}
	\bibinfo{author}{Henkelman, G.} \& \bibinfo{author}{Jonsson, H.}
	\newblock \bibinfo{title}{{Improved tangent estimate in the nudged elastic band
			method for finding minimum energy paths and saddle points}}.
	\newblock \emph{\bibinfo{journal}{J. Chem. Phys.}}
	\textbf{\bibinfo{volume}{113}}, \bibinfo{pages}{9978} (\bibinfo{year}{2000}).
	
\end{thebibliography}

\end{document}